# OPTICAL MANIPULATION OF RELATIVISTIC ELECTRON BEAMS USING THz PULSES*

J. Hebling[#], J. A. Fülöp, M. I. Mechler, L. Pálfalvi, C. Tőke, G. Almási
*Institute of Physics, University of Pécs, 7624 Pécs, Hungary*

*Abstract*

There are implementations and proposals for using microwave or optical radiation for electron acceleration, undulation, deflection, and spatial as well as temporal focusing. Using terahertz (THz) radiation in such applications can be superior to microwave or optical radiation since THz pulses can be generated with significantly smaller temporal jitter to the electron bunch to be manipulated as compared to microwave pulses, and contrary to the optical pulses, the much larger wavelength of THz pulses compares well with typical sizes of electron bunches. Recently generation of ultrashort THz pulses with 1 MV/cm focused electric field was demonstrated, and it is predicted that THz pulses with even 100 MV/cm focused field will be possible. According to the first results of our analytical and numerical studies this field strength is high enough for manipulation of relativistic electron beams by polarised THz pulses in various geometries. It is possible, e.g., to construct a 30 cm long THz undulator for saturated FEL working at 9.6 nm wavelength using electron bunches with 100 MeV energy and 0.42 nC charge.

## INTRODUCTION

In the field of FELs there was a continuous interest in decreasing the working wavelength. Last year the corresponding development resulted in the first X-ray FEL [1], opening the door for a vast of exciting applications. However, because of the simple relation of

$$\lambda_r = \frac{\lambda_u}{2\gamma^2}\left[1 + \frac{K^2}{2}\right], \quad (1)$$

where $\lambda_r$ is the wavelength of the generated radiation, $\lambda_u$ is the period of the undulator, $\gamma$ is the energy of the electrons in the unit of the rest mass, and $K$ is the undulator parameter, using magnet undulators the generation of X-ray radiation by a FEL needs electron energy on the range of 10 GeV. Because of this, the X-ray FEL is a huge and very expensive device.

Nearly three decades ago the microwave undulator was proposed [2] as potentially tunable or short wavelength undulator. However, using microwave radiation for electromagnetic undulator does not result in significantly shorter undulator period than a typical static magnetic undulator. In contrast, according to Eq. 1, using high energy laser radiation with micrometer-scale wavelength range for electromagnetic undulator [3] could result in FEL radiation on the angstrom range using electrons with only a few tens of MeV energy. A further advantage of a laser undulator would be a significantly shorter gain length, since it is proportional to the 5/6 power of the undulator period [4]. In spite of these advantages, as we know, no X-ray FEL (or other FEL) exists with a laser undulator, since the tolerable relative energy spread and the transverse emittance of the electrons decrease with the undulator period (they are proportional to the 1/6 and the 3/2 power of the undulator period, respectively) [5]. Furthermore, the Guoy-phase shift of the focused laser radiation results in a changing effective undulator period.

___________________________________________
*Work supported by Hungarian Research Fund (OTKA), grant numbers 76101 and 78262, and from SROP-4.2.1.B-10/2/KONV-2010-0002
[#]Hebling@fizika.ttk.pte.hu

The latter problem would not be present in the transverse geometry of the laser undulator proposed in Ref. [6]. However, such setup (and similar setups proposed for acceleration [7] or deflection and spatial focusing [8] of electron bunches) would need electron bunches with much smaller transverse size than the wavelength of the laser. Such electron bunches are not available for laser wavelengths of 1 μm (Nd or Yb lasers) or 10 μm ($CO_2$ laser). Contrary to this, by using high intensity THz radiation having a few hundreds of μm wavelength as "laser" undulator or to accelerate, deflect or focus electron bunches requires electron bunches with readily available transverse sizes.

In this paper we would like to demonstrate with a few simple calculations that it could be feasible to use high field THz pulses in transversal geometry for manipulation of relativistic electron bunches. First the state of the art and prospects of high field THz pulse generation will be reviewed. Then the possibility of temporal focusing will be briefly discussed. At the end the feasibility of using a THz undulator with small-scale accelerator for producing XUV and X-ray FEL radiation will be analysed.

## HIGH-FIELD THz-PULSE GENERATION

THz pulses with high energy and high peak electric field are generated either in air-plasma or in nonlinear optical crystals from ultrashort laser pulses. THz pulses with up to 5 μJ energy were generated in plasmas [9]. The frequency content of such pulses can go up to many tens of THz. However, since in the applications considered here lower frequency is needed and the up-scaling of the energy of pulses generated in this way is not obvious we do not discuss this method here.

Using nonlinear optical crystals THz pulses can be generated by second order nonlinear optical processes; optical rectification or difference-frequency generation. In either case fulfilment of phase-matching is needed. Up to now THz pulses with the highest field strength, up to

100 MV/cm, were generated in GaSe from 30 fs long laser pulses utilizing birefringent phase-matching [10]. The highest field was achieved at 30 THz. This frequency is too high, since the corresponding 10 μm wavelength is comparable or shorter than the typical transversal electron bunch size.

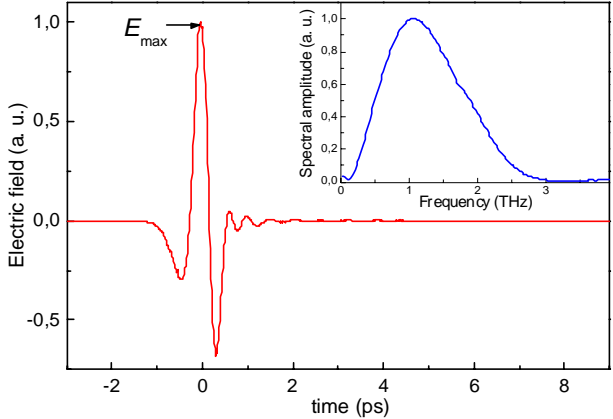

Figure 1: Typical shape of THz pulse generated by TPFP.

THz pulses with the more suitable 0.2 – 1.0 THz mean frequency can be efficiently generated in LiNbO$_3$ [11] using the tilted-pulse-front pumping (TPFP) phase-matching technique [12]. Reducing the THz spot size by imaging 1 MV/cm peak THz field strength was achieved [13]. The typical shape of a THz pulse generated in this way is shown in figure 1. It consists of only one oscillation cycle. The corresponding spectrum (see inset) is very broad with a maximum at about 1 THz. This pulse shape is suitable for temporal focusing, but not suitable for undulation. For the latter purpose THz pulses with a few oscillation cycles are suited. Such pulses can also be efficiently generated by the TPFP technique if specially conditioned pump pulses are used [14, 15].

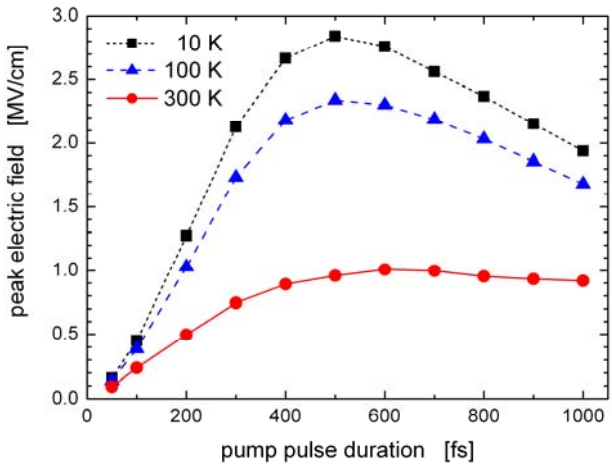

Figure 2: Predicted unfocused peak electric field of THz pulses vs. pump pulse duration for different temperatures.

Up to now the largest energy of THz pulses generated by the TPFP technique is 50 μJ [16]. According to recent calculations [17] using a contact-grating setup [18] for TPFP, generation of THz pulses with tens of mJ energy will be possible using pulses of highly efficient diode-pumped solid-state lasers with optimal pulse duration. Figure 2 depicts the predicted peak electric field strength of the THz pulses generated at different temperatures in LiNbO$_3$ using TPFP versus the pump pulse duration. At low temperatures more than 2 MV/cm field strength is predicted outside of the generating LiNbO$_3$ crystal. Using imaging with strong demagnification or focusing can result in 100 MV/cm field strength in the image/focus plane [17]. The mean frequency is about 0.5 THz, the corresponding wavelength is 600 μm.

According to our calculations, this high field strength and long wavelength (as compared to typical electron bunch transverse sizes) make these pulses suitable for different types of manipulation of relativistic particles. We demonstrate this for two manipulation types, temporal focusing and undulating of electrons.

## TEMPORAL FOCUSING

We re-investigated a recent proposal [19] of Kaplan and Pokrovsky to temporally focus relativistic electrons by a "laser gate," a setting that consists of an electromagnetic (EM) standing wave with polarisation parallel to the propagation direction of the particles, and the interaction length restricted to a fraction of the wavelength. In Ref. 19 analytical calculations were carried out, a laser with 1 μm wavelength was supposed for creating the standing wave, only the effect on the electrons travelling exactly in the plane of one antinode was investigated, and the Coulomb interaction between the electrons was not considered. For a mono-energetic continuous electron beam input, production of a series of electron bunches with as short as 45 zs duration was predicted.

In order to investigate a more realistic situation, we performed particle-particle-interaction (PPI) calculations. Mono-energetic electron bunches with different realistic sizes (from μm to a few tens of μm) was considered. Standing waves created by THz pulses with 0.2 – 2.0 THz frequency and up to 100 MV/cm peak electric field were considered. Figure 3 depicts the calculated electron bunch distribution at five consecutive positions. One position is close to the "laser gate," the other four positions have 5 mm distance increment around the temporal focus (the position at which the bunch is shortest). For this case the Coulomb interaction was switched off and 10 MV/cm peak field was assumed. According to Fig. 3 the duration of the temporally focused electron bunch is a few fs and during the temporal focusing the electron bunch expands in transversal direction, and the shape of the electron bunch distribution become parabolic. The reason of the transversal expansion is that a nonzero magnetic component is present in the EM standing wave field at positions out of the antinode, and the magnitude of this magnetic field is increasing with the distance from the plane of antinode. Since the corresponding Lorentz force is transversal to the travelling direction of the electrons, it does not change their energy. So the electrons at the transversal edges of the electron bunch, having significant transversal velocity, lag behind the electrons at the center resulting in the parabolic distribution. This type of distribution is disadvantageous for applications. However, since the temporal focus is located 3 m away from the "laser gate," there is enough space for placing correcting static magnets, and achieving close-to-flat, 20-nm-long electron bunches seems feasible. Such electron bunches could be used e.g. to generate ultrashort seed pulses for XUV FELs.

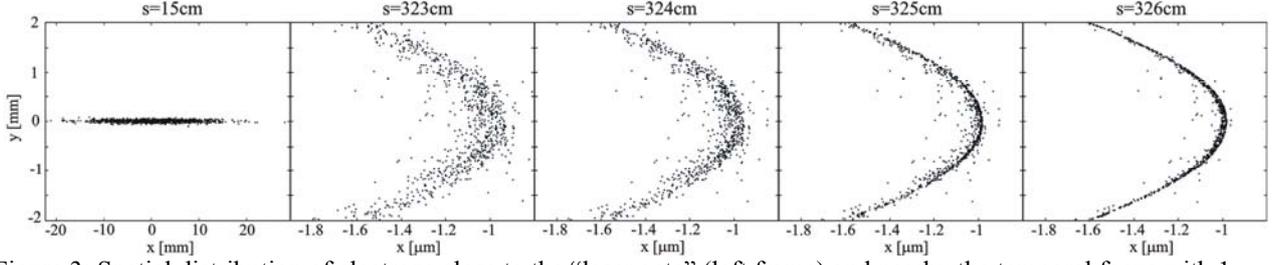

Figure 3: Spatial distribution of electrons close to the "laser gate" (left frame) and nearby the temporal focus with 1 cm flying distance increments (other frames).

We also performed (PPI) calculations for taking into account the Coulomb interaction. According to these preliminary calculations efficient temporal focusing is possible only for bunches with smaller than 10 fC.

## THz UNDULATOR

Here we suggest a device what we call as THz undulator. This is a standing wave created by two counter propagating THz pulses. The relativistic electron bunch propagates at and nearby the plane of the antinode, parallel to it. That is the electrons propagate perpendicular to the propagation directions of the THz pulses. The direction of the electric field of the THz pulses is perpendicular to the propagation direction of the electron bunch. In order to have an impression on the feasibility of an FEL based on such a THz undulator we have calculated the gain length (the length needed for an $e$-times increase of the radiation power) according to

$$L_G = \frac{\sqrt[3]{\lambda_u} \cdot \gamma \cdot \sigma^{2/3} \cdot (1+\Lambda)}{\sqrt[3]{\pi} \cdot \sqrt{3} \cdot K^{2/3}} \cdot \sqrt[3]{\frac{I_A}{I}}, \quad (2)$$

where $\sigma_T$=30 μm is the transverse bunch size, $I_A$=17 kA the Alfven-current, $I$=7.6 kA the peak bunch current (Q = 420 pC the bunch charge and $\sigma_L$=6.6 μm the bunch length), and $\Lambda$=0.4 accounts for different losses. Such parameters are at the limit of current microwave and laser-plasma accelerators. The undulator parameter was calculated, in analogy with the undulator parameter of the planar magnetic undulator, according to

$$K = \frac{\lambda_u \cdot e \cdot E_{THz}}{2\pi \cdot mc^2}, \quad (3)$$

where $\lambda_u = \lambda_{THz}$ is the wavelength of the THz radiation, $e$ the electron charge, $E_{THz}$ the peak electric field of the THz standing wave, $mc^2$ the electron rest energy. In the calculation $\lambda_{THz}$= 600 μm and $E_{THz}$=40 MV/cm values were used. In this case, according to Eq. 3, the undulator parameter is $K = 0.747$.

Figure 4 depicts the necessary electron energy as a function of the generated FEL radiation, and the gain length as the function of the generated FEL radiation calculated from Eq. 1, and Eqs. 2 and 3, respectively, using the parameters given above. The short undulator period of $\lambda_u$ = 600 μm, which is 50 times shorter than a typical undulator period for static magnet undulators, results in a significant decrease in the needed electron energy. For a given FEL radiation wavelength, according to Eqs. 1 and 2, the gain length is proportional to $\lambda_u^{5/6}$. This means that the much shorter undulator period for THz undulator as compared to the period of static magnet undulator results in significantly shorter gain length. It is possible, e.g., to construct a 30 cm long THz undulator for saturated FEL working at 9.6 nm FEL radiation wavelength using electron bunches with only 100 MeV energy. Thus, the sizes of both the undulator and the accelerator are significantly reduced if a THz undulator is used instead of a static magnet undulator.

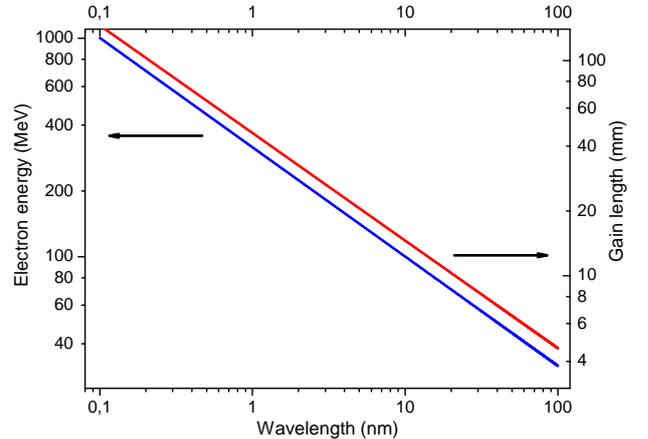

Figure 4: Needed electron energy and gain length vs. FEL radiation wavelength.

Creating THz pulses with 40 MV/cm focused electric field strength will not be a problem in the near future. Supporting such field strength over the needed few times 10 cm length is a challenge. However, using 10 – 20 separate THz pulse sources the elimination of this challenge seems to be feasible.

## CONCLUSIONS

According to our calculations, with the application of THz pulses having 10 MV/cm level peak electric field strength, it is possible to efficiently manipulate relativistic particles. We proposed THz undulator. According to our preliminary calculations, application of THz undulator in FELs seems feasible. Such application makes possible an order of magnitude reduction of the electron energy, and the size of (both of the accelerator and the undulator) FELs. Further numerical calculations are needed to clarify the limitation of such type of applications of THz pulses. For example one has to take into account the effect of space charge both in the case of THz undulator and optical gate. Investigation of more complex arrangements is also necessary, which result in flat electron distribution and allow using higher bunch charge.